\documentclass[a4paper,fleqn]{cas-sc}

\usepackage{caption}
\usepackage{algorithm}
\usepackage{algpseudocode}
\usepackage[authoryear]{natbib}
\usepackage{hyperref}
\hypersetup{colorlinks=true, citecolor=blue, linkcolor=blue, urlcolor=blue}
\usepackage[nameinlink]{cleveref}
\crefname{figure}{fig.}{figures}
\Crefname{figure}{Fig.}{Figures}
\crefname{Table}
\Crefname{Table}
\usepackage{booktabs}
\usepackage[table]{xcolor}
\usepackage{multirow}
\usepackage{geometry}
\usepackage{adjustbox} 
\usepackage{amsmath}
\usepackage{amssymb}
\usepackage{enumitem}
\usepackage[T1]{fontenc}

\def\tsc#1{\csdef{#1}{\textsc{\lowercase{#1}}\xspace}}
\tsc{WGM}
\tsc{QE}

\begin{document}
\let\printorcid\relax
\let\WriteBookmarks\relax
\def\floatpagepagefraction{1}
\def\textpagefraction{.001}
\shorttitle{DisenReason}
\shortauthors{xx et~al.}

\title [mode = title]{DisenReason: Behavior Disentanglement and Latent Reasoning for Shared-Account Sequential Recommendation}  



%

\author[1]{Jiawei Cheng}



\ead{chengjiawei@stu.cqu.edu.cn}


\credit{Conceptualization of this study, Methodology, Software, Writing}

\affiliation[1]{organization={School of Big Data and Software Engineering},
            addressline={Chongqing University}, 
            city={Chongqing},
            postcode={401331}, 
            country={China}}

\author[1]{Min Gao}
\ead{gaomin@cqu.edu.cn}
\ead[url]{}
\credit{Supervision, Writing}
\cormark[1]
\cortext[1]{Corresponding author}

\author[1]{Zongwei Wang}
\ead{zongwei@cqu.edu.cn}
\credit{Supervision, Writing}

\author[2]{Xiaofei Zhu}
\ead{zxf@cqut.edu.cn}
\credit{Supervision}
\affiliation[2]{organization={College of Computer Science and Engineering},
            addressline={Chongqing University of Technology}, 
            city={Chongqing},
            postcode={401135}, 
            country={China}}

\author[1]{Zhiyi Liu}
\ead{liuzhiyi@stu.cqu.edu.cn}
\credit{Visualization}

\author[3]{Wentao Li}
\ead{wl226@leicester.ac.uk}
\credit{Supervision}
\affiliation[3]{organization={ School of Computing and Mathematical Sciences},
            addressline={University of Leicester}, 
            city={Leicester},
            postcode={401135}, 
            country={United Kingdom}}

\author[4]{Wei Li}
\ead{liwei00128@cqu.edu.cn}
\credit{Supervision}
\affiliation[4]{organization={Department of Plastic Surgery},
            addressline={Chongqing University Central Hospital, Chongqing Emergency Medical Center}, 
            city={Chongqing},
            postcode={400000}, 
            country={China}}

\author[5]{Huan Wu }
\ead{wuhuan@tongji.edu.cn}
\credit{Supervision}
\affiliation[5]{organization={College of Environmental Science and Engineering},
            addressline={Tongji University}, 
            city={Shanghai},
            postcode={200092}, 
            country={China}}



\begin{abstract}
Shared-account usage is common on streaming and e-commerce platforms, where multiple users share one account. Existing shared-account sequential recommendation (SSR) methods often assume a fixed number of latent users per account, limiting their ability to adapt to diverse sharing patterns and reducing recommendation accuracy. Recent latent reasoning technique applied in sequential recommendation (SR) generate intermediate embeddings from the user embedding (e.g, last item embedding) to uncover users' potential interests, which inspires us to treat the problem of inferring the number of latent users as generating a series of intermediate embeddings, shifting from inferring preferences behind user to inferring the users behind account. However, the last item cannot be directly used for reasoning in SSR, as it can only represent the behavior of the most recent latent user, rather than the collective behavior of the entire account. To address this, we propose DisenReason, a two-stage reasoning method tailored to SSR. DisenReason combines behavior disentanglement stage from frequency-domain perspective to create a collective and unified account behavior representation, which serves as a pivot for latent user reasoning stage to infer the number of users behind the account. Experiments on four benchmark datasets show that DisenReason consistently outperforms all state-of-the-art baselines across four benchmark datasets, achieving relative improvements of up to 12.56\% in MRR@5 and 6.06\% in Recall@20.

\end{abstract}





\maketitle

\section{Introduction}\label{}
In many digital services such as streaming media, e-commerce, online learning, and smart home platforms, users constantly interact with systems over time, forming behavior sequences that reveal their evolving interests and needs. Sequential recommendation~\citep{peintner2025hypergraph,liu2024selfgnn, wang2024unveiling,wang2025id} has emerged as an effective solution for predicting users’ next interactions by leveraging these temporal dynamics. However, a widely adopted assumption underlying existing sequential recommenders is that each account is operated by a single individual, which often breaks in real-world applications where shared-account usage is common. For example, family members may jointly use a single VIP video account, music subscription account, or online shopping account. To address this prevalent practical scenario, the shared-account sequential recommendation (SSR) task~\citep{ma2019pi} was proposed, which targets accurate recommendations when multiple real users operate the same account.

To tackle the SSR task, existing studies have explored extensions of sequential recommendation models by leveraging recurrent architectures to encode temporal dependencies~\citep{guo2022time,guo2022reinforcement} or graph neural networks to propagate collaborative signals~\citep{zhang2025lightweight}. These methods assume that each shared account consists of a fixed number of latent users and then attempt to disentangle their sequential behaviors accordingly. Although this assumption simplifies model design, it is misaligned with real usage patterns. Shared accounts in practice exhibit substantial diversity in the number of underlying users, ranging from two siblings to several household members. Enforcing a uniform and predetermined user count limits the model’s ability to adapt to different sharing scenarios. Consequently, the key challenge remains unresolved in the SSR task: how to accurately infer and represent the dynamic number of latent users behind a shared account to support high-quality recommendations.

\begin{figure*} 
  \centering
  \includegraphics[width=\linewidth]{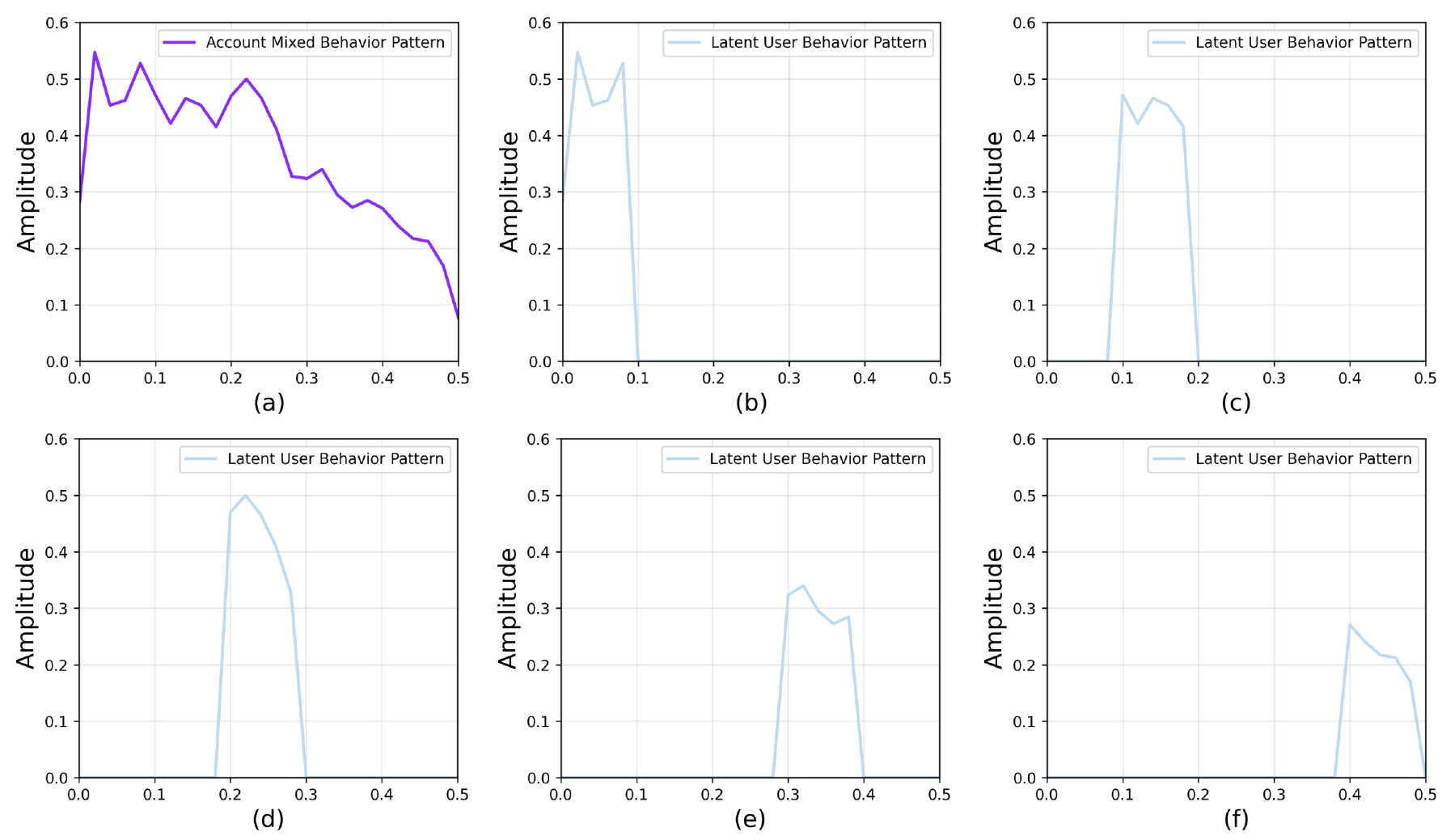}
  \caption{Visualization of frequency-domain disentanglement for shared-account behavior sequence.}
  \label{fig:ex}
\end{figure*}

Recent latent reasoning paradigms~\citep{tang2025think} have achieved strong performance in sequential recommendation by enabling models to conduct implicit multi-step inference based on last item embedding. 
Unlike text-based reasoning, latent reasoning occurs within the latent space of user behavior, where intermediate results emerge as hidden embeddings represented diverse interests rather than textual. 
This paradigm inspires us that it could be a promising direction to transform the problem of learning the number of latent users in an account into a process of generating intermediate results in the latent space, which represents the behavior space of the account. 

Yet, existing latent reasoning methods in sequential recommendation are not directly applicable to shared-account scenarios. Traditional sequential recommendation assumes that each account corresponds to a single individual. Under this condition, it is natural to use the last item as the representation of the user's current intent and as the starting pivot for reasoning. In shared-account scenarios, however, this assumption no longer holds. The interaction sequence becomes a mixture of behaviors from multiple underlying users. While the last item may reflect the behavior of one user at a specific moment, it fails to capture the collective nature of the shared account. Without a unified pivot that represents the account as a whole, latent reasoning methods struggle to uncover the hidden users behind the account. This limitation calls for a global behavioral pivot from this mixed interaction sequence that captures the collective dynamics.


We address this challenge by adopting a temporal-signal perspective, viewing shared-account behavior as a mixture of multiple temporal patterns. Signal processing theory~\citep{pitas2000digital,acharya2005image} suggests that an entangled signal can be decomposed into sinusoidal components of distinct frequencies, where low frequencies represent stable, long-term sources, while high frequencies represent unstable, short-term sources. Analogously, we posit that different latent users within a shared account also exhibit distinct temporal rhythms: some users dominate stable, long-term preferences (low-frequency), while others contribute to bursty, short-term interactions (high-frequency).
This insight motivates us to examine the frequency structures embedded in shared account sequences. Our empirical analysis validates this idea. As illustrated in ~\Cref{fig:ex}, the mixed interaction sequence exhibits a broad and continuous energy distribution. We then disentangle this mixed distribution into equal-width intervals along the frequency axis, obtaining five distinct frequency components. These disentangled components each exhibit concentrated energy within their respective frequency bands, with minimal overlap. This observation suggests that we can capture different behavior patterns in the mixed sequence from a frequency-domain perspective. By combining these components, we reconstruct a complete representation of the account's collective behavior, thereby establishing a pivot for latent user reasoning.

Building on this insight, we propose a novel two-stage reasoning framework named \textbf{DisenReason}, which aims to obtain the account-level representation for reasoning pivot and uncover the number of user behind account via latent reasoning on this pivot. In stage one, DisenReason reinterprets the mixed interaction sequence in the frequency domain using Fast Fourier Transform (FFT)~\citep{brigham1988fast} and disentangles it into different behavioral patterns. Since these patterns contribute unequally to the shared account, we introduce a Mixture-of-Experts~\citep{chen2022moe} fusion scheme to adaptively weight and integrate them into a reliable account-level representation, which serves as the pivot for subsequent reasoning. In stage two, DisenReason conducts progressive residual reasoning to sequentially uncover latent users based on pivot. At each step, an intermediate embedding is generated from the pivot to represent an identified user, followed by a residual technique that removes this inferred user from the pivot to avoid redundant reasoning. The reasoning loop continues until the semantic similarity between newly inferred and previously identified users surpasses a preset threshold, signaling that all latent users behind the shared account have been successfully recovered. Experiments on four benchmark datasets demonstrate that our proposed DisenReason outperforms previous state-of-the-art baselines. 

The main contributions of this paper are summarized as follows:
\begin{itemize}[label=\textbullet]
    \item To the best of our knowledge, this is the first work to introduce latent reasoning into the Shared-Account Sequential Recommendation task. This novel approach aims to tackle the challenge of accurately identifying the hidden users in shared account scenarios.

    \item  We propose DisenReason, a novel two-stage reasoning framework that integrates behavior disentanglement to derive account behavior representation and subsequently performing latent reasoning over them, aiming to infer the number of latent users behind the account.
    
    \item Extensive experiments conducted on four benchmark datasets demonstrate that DisenReason consistently outperforms state-of-the-art methods across various shared-account scenarios, confirming its effectiveness and robustness.
\end{itemize}

\section{Related Work}
\subsection{Sequential Recommendation}
Sequential Recommendation (SR) aims to predict a user’s next interaction by modeling the transition dependencies among items within a behavioral sequence. Early approaches rely on Markov Chain-based models~\citep{he2016fusing,rendle2010factorizing} to alleviate data sparsity issues. However, these methods are limited to capture the dynamics of user preferences. With the rapid progress of deep learning, various neural sequential recommenders~\citep{peintner2025hypergraph,liu2024selfgnn, li2026multi, jiang2026hierarchical, wang2026graph} are introduced to learn more expressive sequential representations. Subsequently, self-attention mechanism~\citep{vaswani2017attention} demonstrate remarkable effectiveness in modeling long-range dependencies in sequential data, inspiring a series of transformer-based SR models~\citep{kang2018self,sun2019bert4rec,deng2025heterrec,pan2026multi, liu2025multi}. Despite their success, most of these models presume a one-to-one correspondence between accounts and users, making them unsuitable for shared-account scenarios.

Building upon SR research, Shared-Account Sequential Recommendation (SSR) extends the task to scenarios where multiple users share a single account~\citep{verstrepen2015top,duan2025user,wang2025user}. The key objective is to capture the heterogeneous preferences of latent users and generate personalized recommendations.  Early SSR studies mainly employed RNN-based frameworks~\citep{ma2019pi,sun2021parallel} to infer user-specific patterns within shared accounts. However, these models often suffer from gradient vanishing and fail to capture long-range dependencies. To address these shortcomings, subsequent works~\citep{guo2021gcn,guo2022time} introduced graph-based SSR architectures augmented with attention mechanisms, enabling the propagation of user-dependent signals through graph structures. Other work~\citep{guo2022reinforcement} uses reinforcement learning-based method for SSR. While such methods have achieved remarkable improvements, they typically assume a fixed number of latent users for all accounts and learn directly from mixed interaction sequences, which limits their ability to disentangle user behaviors effectively and ultimately constrains recommendation accuracy.

In contrast, our proposed two-stage reasoning method adaptively learns the correct number of latent users within each shared account in a progressive reasoning manner, thereby enabling more accurate and personalized recommendations.

\subsection{Reasoning-based Recommendation}
Inspired by LLMs' Chain of Thought~\citep{wei2022chain,geiping2025scaling,chen2025language}, \cite{tang2025think} introduces the ``think before acting'' paradigm into sequential recommendation, enhancing the computational capability of sequential recommendation through implicit reasoning. To achieve this, they proposed two reasoning learning approaches: Ensemble Reasoning Learning (ERL) and Progressive Reasoning Learning (PRL). The former leverages the idea of ensemble learning to construct multi-order user representations, aiming to comprehensively capture latent interest distributions from diverse perspectives. The core idea of the latter is to design a progressive distribution sharpening strategy to guide the intermediate reasoning chains, gradually approximating the user’s true preference distribution.  Subsequently, PLR~\citep{tang2026plr} introduces width-level computational scaling to sequential recommendation, addressing the diminishing returns of depth-only latent reasoning by exploring multiple reasoning trajectories simultaneously. It employs learnable trigger tokens to construct parallel streams in latent space, preserves diversity via global regularization, and adaptively aggregates multi-stream outputs through a mixture-of-reasoning-streams mechanism.

In this paper, we apply this technique to SSR. Unlike directly applying the technique to SSR, we have carefully designed it to better suit the SSR task. These designs include learning an appropriate reasoning pivot as the starting point for reasoning, rather than directly using the last item as the reasoning starting point as in~\citep{tang2025think}, and incorporating a residual mechanism during the reasoning process to prevent redundant reasoning.

\subsection{Fourier Transform in Recommendation}
Fourier transform has been an important tool in digital signal processing and graph signal processing for decades~\citep{pitas2000digital,acharya2005image,peng2022less,cheung2020graph}. There are a variety of works that incorporate fourier transform in computer vision~\citep{rao2021global,xu2020learning,suvorov2022resolution,Zhao24CDH,Zhang25DXPZW} and natural language processing~\citep{tamkin2020language,ai2025revisit}. Very recent works try to leverage fourier transform enhanced model for long-term series forecasting~\citep{zhou2022fedformer,wu2025affirm} and sequential recommendation. In sequential recommendation, FMLP-Rec~\citep{zhou2022filter} first introduced frequency-based filtering using MLPs to uncover periodic patterns in user-item interactions. FEARec~\citep{du2023frequency} proposed advanced architectures with frequency-aware attention mechanisms. Additionally, some studies~\citep{qiu2022contrastive,yang2024adaptive} have incorporated fourier-based data augmentation for contrastive learning, further enhancing representation robustness. BSARec~\citep{shin2024attentive} introduced fine-grained frequency adjustment, aiming to better capture subtle sequential signals. MUFFIN~\citep{baek2025muffin} adopts a dual filtering architecture, comprising GFM and LFM, to exploit the full frequency spectrum. 


Despite these advances, existing approaches primarily focus on single-user scenarios, where the entire interaction sequence is assumed to originate from one individual. In contrast, our work leverages fourier transform to disentangle the mixed behavioral signals in shared-account sequences, enabling the model to separate different users’ patterns from a frequency perspective. This novel application of frequency analysis opens up new possibilities for handling multi-user accounts in recommendation systems.

\begin{table}[ht]
\centering
\caption{Notations used in DisenReason}
\label{tab:notations}
\begin{tabular}{ll}
\toprule
\textbf{Notation} & \textbf{Definition} \\
\midrule
\( \mathcal{A} \) & Set of shared accounts \\
\( A_j \) & The \(j\)-th shared account \\
\( V \) & Set of items \\
\( v_i \) & The \(i\)-th item \\
\( \mathcal{M} \) & Interaction matrix where \(\mathcal{M}_{ij}=1\) indicates account \(A_i\) interacted with item \(v_j\) \\
\( S_k = [v_1^k, v_2^k, \dots, v_{s}^k] \) & Interaction sequence of account \(A_k\) in chronological order \\
\( m \) & Number of accounts \\
\( n \) & Number of items \\
\( d \) & Embedding dimension \\
\( \mathbf{E}_A, \mathbf{E}_V \) & Initial account and item embeddings from ID lookup \\
\( \mathbf{H}_A, \mathbf{H}_V \) & Final account and item embeddings after LightGCN propagation \\
\( \mathbf{H}_{\text{seq}} \) & Sequence representation of an account (item embeddings looked up from \(\mathbf{H}_V\)) \\
\( \mathcal{F}, \mathcal{F}^{-1} \) & Fast Fourier Transform and its inverse \\
\( \mathbf{F}_{\text{seq}} \) & Frequency-domain representation of the sequence, \( f = \lceil s/2 \rceil \) \\
\( B \) & Bandwidth for frequency decomposition \\
\( Z = \lceil f/B \rceil \) & Number of frequency sub-bands \\
\( \mathbf{F}_{\text{seq}}^{(z)} \) & The \(z\)-th frequency sub-band, \( \mathbb{R}^{B \times d} \) \\
\( \tilde{\mathbf{H}}_{\text{seq}}^{(z)} \) & Time-domain representation of the \(z\)-th behavioral pattern (after IFFT and padding) \\
\( \mathbf{w}_f \) & Adaptive fusion weights for behavioral patterns \\
\( \bar{\mathbf{H}}_{\text{seq}} \) & Fused multi-behavior sequence representation \\
\( \bar{\mathbf{h}}_{seq} = \bar{\mathbf{H}}_{\text{seq}}[-1] \) & Reasoning pivot (last hidden state of fused sequence) \\
\( \mathbf{r}_p \) & Learnable reasoning position embedding \\
\( \mathbf{r}^{(t)} \) & Reasoning state at step \(t\) (initialized as \(\mathbf{r}^{(0)} = \bar{\mathbf{h}}_{seq} + \mathbf{r}_p\)) \\
\( \phi(\cdot) \) & Reasoning function (shared with stage-one disentanglement) \\
\( \mathbf{u}^{(t)} \) & Inferred representation of the \(t\)-th latent user \\
\( T \) & Total number of inferred latent users for the account \\
\( \alpha \) & Similarity threshold for terminating progressive reasoning \\
\( \mathbf{h}_{\text{final}} \) & Aggregated account representation from all inferred users \\
\( \mathbf{W}_A, \mathbf{b}_A \) & Transformation matrix and bias for final prediction \\
\( \mathcal{L}_{\text{rec}} \) & Cross-entropy loss for recommendation \\
\( \mathcal{L}_{\text{aux}} \) & Auxiliary recommendation loss from intermediate user representations \\
\( \beta \) & Weight for auxiliary loss \\
\( L \) & Number of LightGCN layers \\
\bottomrule
\end{tabular}
\end{table}

\section{Research Objective}

We structure our research objective by first identifying the limitations in existing studies, then formulating the core problems to be solved, and finally summarizing our key contributions.

\textbf{Limitations of Existing Research.} Existing SSR methods typically assume a fixed number of latent users per account, failing to reflect real-world sharing scenarios where user counts vary dynamically. Moreover, conventional latent reasoning paradigms rely on the last item as the reasoning pivot, which captures only the most recent user and neglects the collective behavior of the entire account.

\textbf{Problems to be Addressed.} We address two key challenges: (1) how to construct a unified account-level representation that captures the collective behavior of shared account; and (2) how to adaptively infer the variable number of latent users behind each account without a predetermined count.

\textbf{Contributions of Our Research.} We propose DisenReason, a two-stage reasoning framework for SSR. It introduces (1) a behavior disentanglement module in stage one that disentangles mixed sequence into distinct behavioral patterns to construct a reliable account pivot; (2) a progressive residual reasoning mechanism in stage two that iteratively infers latent user and removes it from the pivot to avoid redundancy of reasoning; (3) Extensive experiments show that DisenReason consistently outperforms state-of-the-art methods.

\section{Preliminaries}

\subsection{Problem Formulation}

Let $V=\{v_1, v_2, .., v_i, .., v_n\}$ is the set of items, where $v_i$ denotes the $i$-th item. The set of shared accounts is denoted as $\mathcal{A}=\{A_1, A_2, ..,A_k,.., A_m\}$, where $A_k$ represents the $k$-th shared account. Moreover, the set of sequences is denoted as $\mathcal{S}=\{S_1, S_2, .., S_k, .., S_n\}$, where $S_k \in \{v_1^{k}, v_2^{k}, .., v_s^{k}\}$ is the hybrid sequence of shared account $A_k$. 
Given $S_k$ and $A_k$, the task of SSR is recommend the next item $v_{s+1}^{k}$ that $A_k$ is most likely to interact, based on the account's hybrid sequence $S_k$. The probabilities of item $v_{s+1}^{k}$ is defined as:
\begin{equation}
P(v_{s+1}^{k} \mid S_k, A_k) \sim f(S_k, A_k),
\end{equation}
where $P(v_{s+1}^{k} \mid S_k, A_k)$ denotes the probability of recommending $v_{s+1}^{k}$ to $A_k$ given its historical hybrid sequence $S_k$, and $f(S_k, A_k)$ is the function designed to estimate the probability.

\begin{figure*}
  \centering
  \includegraphics[width=\linewidth]{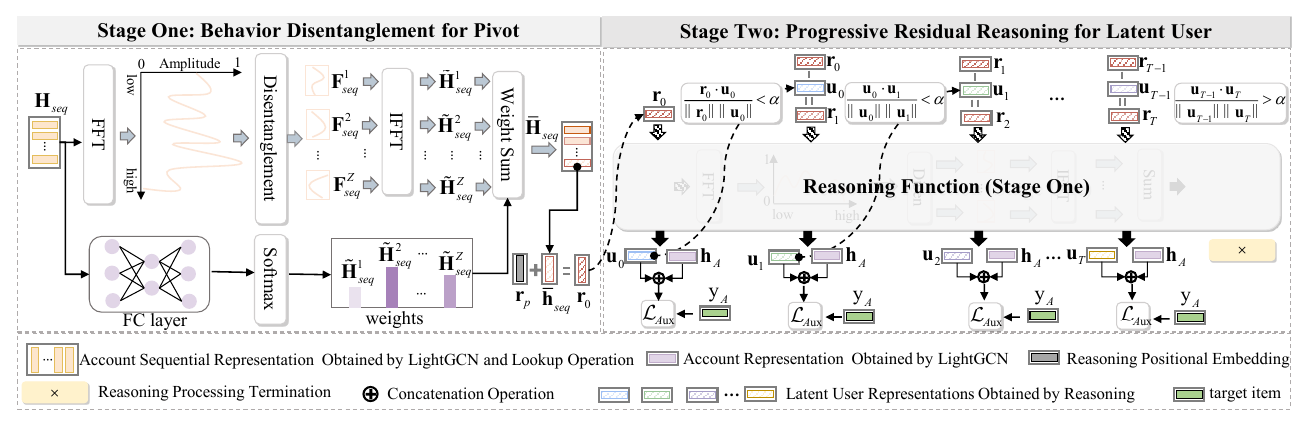}
  \caption{Overview of the proposed DisenReason framework. This figure shows the details of DisenReason: stage one focuses on disentangling the mixed behavioral sequence to create a unified account pivot, while stage two progressively infers latent users based on this pivot.}
  \label{fig:frame}
\end{figure*}

\subsection{Fourier Transform}\label{sec:3.2}
Discrete Fourier Transform (DFT) is a popular computing method for data analysis, signal processing and machine learning. Given a finite sequence $\{ {x_i}\} _{i = 1}^n$, the DFT convert the original sequence into the frequency signals in the frequency domain by:
\begin{equation}
\label{eq:1}
    {X_k} = \sum\limits_{i = 1}^n {{x_i}W_n^{ik}} ,{\text{1}} \leqslant {\text{k}} \leqslant {\text{n}},
\end{equation}
where $n$ represents the sequence length, $W_n^{ik}$ is the rotation factor, ${X_k}$ is a complex number represents the signal with frequency ${\omega _k} = 2\pi k/n$. Therefore, we can decompose a series of values into different frequency components. It is noteworthy that the DFT is a one-to-one unique mapping operation between the time domain and the frequency domain. The frequency representation 
${X_k}$ can be transformed back into the time domain representation through the Inverse Discrete Fourier Transform (IDFT), which is formulated as:
\begin{equation}
    {x_i} = \frac{1}{n}\sum\limits_{k = 1}^n {{X_i}W_n^{ - ik}},
\end{equation}
For a real input $x_i$, it has been proven that its DFT is conjugate symmetric, i.e., ${X_k} = X_{n - k}^*$, where $*$ denotes the conjugate operation. This indicates that half of the DFT contains the complete frequency characteristics. If we perform Inverse Discrete Fourier Transform (IDFT) on $\{ {X_k}\} _{k = 1}^{\left\lceil {n/2} \right\rceil +1 }$, a real signal can be recovered. The Fast Fourier Transform (FFT)
algorithm is a fast algorithm for computing the DFT, reducing the complexity to compute DFT from $\mathcal{O}(N^2)$ to $\mathcal{O}(NlogN)$. The Inverse FFT (IFFT), which has a similar form to the DFT. In this paper, we denote FFT and IFFT by $\mathcal{F}$ and $\mathcal{F}^{-1}$, respectively

\begin{algorithm}[t]
\caption{Forward Pass of DisenReason}
\label{al:1}
\begin{algorithmic}[1]
\Require Shared account $A_k$ with interaction sequence $S_k = [v^{k}_1, v^{k}_2, \ldots, v^{k}_s]$; jointly trained LightGCN embeddings $\mathbf{H}_\mathcal{A}$ and $\mathbf{H}_V$.
\Ensure Predict next item probability $P(v^{k}_{s+1} \mid S_k, A_k)$.

\Statex \textbf{Stage One: Behavior Disentanglement for Pivot}

\State Obtain sequence representation $\mathbf{H}_{seq}$ by looking up $\mathbf{H}_V$ according to $S_k$.
\State Compute FFT: $\mathbf{F}_{seq} = \mathcal{F}(\mathbf{H}_{seq})$, where $f = \lceil s/2 \rceil$.
\State Partition frequency axis into $Z = \lceil f/B \rceil$ sub-bands of equal width $B$:
   $\mathbf{F}_{seq}^{(z)} = \mathbf{F}_{seq}[b_z:b_{z+1}]$, $b_z = z \cdot B$, $z=1,\ldots,Z$.
\For{each sub-band $z$}
    \State Pad $\mathbf{F}_{seq}^{(z)}$ with zeros to length $f$ and apply IFFT: $\tilde{\mathbf{H}}_{seq}^{(z)} = \mathcal{F}^{-1}(ZeroPad(\mathbf{F}_{seq}^{(z)}))$.
\EndFor
\State Compute adaptive fusion weights: $W_f = Softmax(FC(\mathbf{H}_{seq}))$.
\State Obtain fused sequence: $\bar{\mathbf{H}}_{seq} = \sum_{z=1}^{Z} w_{f,z} \, \tilde{\mathbf{H}}_{seq}^{(z)}$.
\State Extract reasoning pivot: $\mathbf{p} = \bar{\mathbf{H}}_{seq}[-1]$ (last hidden state).

\Statex \textbf{Stage Two: Progressive Residual Reasoning for Latent Users}

\State Initialize reasoning state: $\mathbf{r}^{(0)} = \mathbf{p} + \mathbf{r}_p$, where $\mathbf{r}_p$ is a learnable reasoning position embedding.
\State Set $t = 1$.
\Repeat
    \State Infer $t$-th latent user: $\mathbf{u}^{(t)} = \phi(\mathbf{r}^{(t-1)})$, where $\phi$ is the reasoning function (implemented as the same disentanglement module applied to a single vector).
    \State Residual update: $\mathbf{r}^{(t)} = \mathbf{r}^{(t-1)} - \mathbf{u}^{(t)}$.
    \If{$t \geq 2$ and $cosine\_similarity(\mathbf{u}^{(t)}, \mathbf{u}^{(t-1)}) > \alpha$}
        \State \textbf{break} \Comment{Terminate when consecutive users become too similar}
    \EndIf
    \State $t = t + 1$.
\Until{termination condition met}
\State $T = t$ (number of inferred users).
\State Aggregate user representations: $\mathbf{h}_{final} = \frac{1}{T} \sum_{t=1}^{T} \mathbf{u}^{(t)}$.

\Statex \textbf{Prediction}

\State Compute next item probability:
   $P(v^{k}_{s+1} \mid S_k, A_k) = \text{softmax}\big( \mathbf{W}_A \cdot [\mathbf{h}_{final}; \mathbf{h}_A] + \mathbf{b}_A \big)$.
\end{algorithmic}
\end{algorithm}

\section{Proposed Framework: DisenReason}
In this section, we present DisenReason, which is capable of adaptively learning the number of latent users within a shared account through self-reasoning. As illustrated in~\Cref{fig:frame}, the overall framework of DisenReason has two stages: stage one aims to learn the reasoning pivot and stage two uses the reasoning pivot to infer the number of latent users in a progressive reasoning manner. Key symbols and their meanings
are summarised in~\Cref{tab:notations}. The pseudocode for DisenReason is shown in Algorithm~\ref{al:1}. The detailed formulation and design of DisenReason is provided  below. 

\subsection{Stage One: Behavior Disentanglement Reasoning for Pivot}
The effectiveness of downstream reasoning is inherently tied to the quality of input representations. In our framework, the initial account and item embeddings are obtained via a standard embedding layer that maps each ID to a $d$-dimensional vector. Such ID-based embeddings are semantically weak, as they encode only identity information without any collaborative context. Feeding these raw embeddings directly into the subsequent two-stage reasoning process would hinder the model's ability to accurately infer latent users. To address this, we enrich the initial embeddings by incorporating global collaborative signals before entering the core reasoning stages.

Specifically, we construct the interaction matrix $\mathcal{M} \in \mathbb{R}^{m \times n}$, where $\mathcal{M}_{ij} = 1$ indicates that account $A_i$ interacted with item $v_j$. Let $\mathbf{E}_\mathcal{A} \in \mathbb{R}^{m \times d}$ and $\mathbf{E}_V \in \mathbb{R}^{n \times d}$ denote the initial embedding matrices for accounts and items, respectively. Given $\mathbf{E}_\mathcal{A}$, $\mathbf{E}_V$, and $\mathcal{M}$, we adopt LightGCN~\citep{he2020lightgcn} to propagate collaborative signals across the global interaction graph. LightGCN is a lightweight graph convolutional network that simplifies traditional GCN by removing nonlinear transformations and feature transformations, while retaining the essential neighborhood aggregation operation. The graph convolution can be formulated as follows:
\begin{align}
\textbf{E}_\mathcal{A}^{(l+1)} &= \tilde{\mathcal{M}} \textbf{E}_V^{(l)}, \;
\textbf{E}_V^{(l+1)} = \textbf{E}_\mathcal{A}^{(l)}\tilde{\mathcal{M}}. \\
\tilde{\mathcal{M}} &= D^{-\frac{1}{2}} \mathcal{M} D^{-\frac{1}{2}},
\end{align}
where $D \in \mathbb{R}^{m \times n}$ is the degree matrix. After applying $L$ layers of graph convolution, the embeddings of each node obtained from all layers 
are aggregated through average pooling to derive the final representations of accounts and items. 
Formally, the final embeddings are defined as:
\begin{equation}
\textbf{H}_\mathcal{A} = \frac{1}{L} \sum_{l=1}^{L} \textbf{E}_\mathcal{A}^{(l)}, 
\textbf{H}_V = \frac{1}{L} \sum_{l=1}^{L} \textbf{E}_V^{(l)}, 
\label{Eq.6}
\end{equation}
where $\textbf{H}_\mathcal{A} \in \mathbb{R}^{m \times d}$ and $\textbf{H}_V \in \mathbb{R}^{n \times d}$. $\textbf{E}_\mathcal{A}^{(l)}$ and $\textbf{E}_V^{(l)}$ denotes the account and item representations at the $l$-th layer.
Based on the item representations matrix $\textbf{H}_V$, we can obtain the account sequence representation $\textbf{S}_\mathcal{A}$ according to input account's interaction sequence $S_A \in \{v_1, v_2, \cdots, v_s\}$, which will be used for the subsequent reasoning process:
\begin{equation}
\begin{array}{c}
\mathbf{H}_{seq} = Lookup(\mathbf{H}_V, S_A),
\end{array}
\end{equation}
where $\textbf{H}_{seq} \in \mathbb{R}^{s \times d}$ is the account sequence representation and $s$ is the sequence length. The $lookup$ operation retrieves the corresponding item embeddings and aligns them temporally to form the account’s behavioral sequence representation.

As established in the introduction, constructing an account-level pivot is essential for the reasoning process, as it directly enables the subsequent stage to accurately infer latent users. To this end, in stage one, we aim to derive such a pivot that captures the collective behavior of the entire account, serving as the basis for reasoning over the account to uncover its underlying users.

Specifically, we first transform the entangled sequence representation $\mathbf{H}_{seq}$ into the frequency domain via, which allows us to disentangle the mixed signals from a frequency perspective, thereby obtaining relatively independent representations of latent behavioral patterns. The operation is denoted as:
\begin{equation}
    \mathbf{F}_{seq} = \mathcal{F}(\mathbf{H}_{seq}),
\end{equation}
where $\mathcal{F}$ denotes the Fast Fourier Transform (FFT) operation (Section~\ref{sec:3.2}). Due to the conjugate symmetry property of FFT for real-valued inputs, only half of the frequency components need to be retained. Thus, the transformed representation is $\mathbf{F}_{seq} \in \mathbb{R}^{f \times d}$, where $f = \lceil s/2 \rceil$ corresponds to the number of unique frequency bins.

To disentangle mixed behavioral patterns, we partition $\mathbf{F}_{seq}$ along the frequency axis into $Z$ sub-bands of equal width $B$. Specifically, the $z$-th sub-band is obtained by slicing the frequency components within the range $[b_z, b_{z+1}]$, where $b_z = z \cdot B$, and the total number of sub-bands is $Z = \lceil f / B \rceil$. Each sub-band $\mathbf{F}_{seq}^{(z)} \in \mathbb{R}^{B \times d}$ captures a specific frequency interval, corresponding to distinct behavioral rhythm:
\begin{align}
\mathbf{F}_{seq}^{(z)} &= \mathbf{F}_{seq}[b_z : b_{z+1}], \; 
b_z = z \cdot B. 
\end{align}
The resulting frequency-domain representation $\mathbf{F}_{seq}^{(z)}$ contains complex-valued components and cannot be directly utilized. To obtain usable representations, we apply the IFFT to convert each sub-band back to the time domain. This yields a set of temporal representations, each reflecting a distinct behavioral rhythm exhibited by the shared account.
The process can be formulated as follows:
\begin{align}
\tilde{\textbf{H}}_{seq}^{(z)} &= \mathcal{F}^{-1}(Padding(\mathbf{F}_{seq}^{(z)})),
\end{align}
where $\tilde{\textbf{H}}_{seq}^{(z)} \in \mathbb{R}^{s \times d}$ indicates the representation of $z$-th behavior pattern.
$Padding$ means zero-padding, which extends the frequency length of $\mathbf{F}_{seq}^{(z)}$ to match that of $\mathbf{F}_{seq}$, so that the inverse transformation can be performed.
However, not all frequency components contribute equally to the overall account representation. Among these components, some capture the dominant behavioral patterns of the shared account, while others primarily reflect noise or transient interactions that carry limited semantic significance. To adaptively determine the contribution of each behavioral pattern, 
we learn the relative importance of different behavioral patterns and employ weight sum. The operation is defined as follows:
\begin{align}
W_f &= Softmax(FC(\mathbf{H}_{seq})), \\
\bar{\textbf{H}}_{seq} &= \sum\limits_{z = 1}^{Z} {{w^z_f}\tilde {\textbf{H}}_{seq}^{(z)}},
\end{align}
where $FC(\cdot)$ denotes a fully connected layer that learns the weight $w_f^z \in W_f$ for each behavioral pattern from the input sequence $\mathbf{H}_{seq}$. This adaptive fusion mechanism emphasizes the most informative behavioral rhythms while suppressing redundant or noisy components, yielding a comprehensive multi-behavior account sequential representation $\bar{\mathbf{H}}_{seq}$. Finally, we take the last hidden state $\bar{\mathbf{h}}_{seq} = \bar{\mathbf{H}}_{seq}[-1]$ as the aggregated account-level behavioral signal, which serves as the reasoning pivot for the subsequent progressive reasoning stage.

\subsection{Stage Two: Progressive Residual Reasoning for Latent User}
In this stage, our goal is to leverage the reasoning pivot to initiate progressive residual reasoning, thereby implicitly recovering the latent users behind the shared account. Since the reasoning function $\phi(\cdot)$ in stage two reuses the same disentanglement operation from stage one, without proper distinction, the model would fail to differentiate between the two stages, leading to identical processing and undermining the reasoning process. 
To address the potential confusion caused by reusing the same operation, we introduce a reasoning positional embedding $\textbf{r}_p$ that explicitly distinguishes stage two from stage one. This embedding is combined with the account pivot $\bar{\textbf{h}}_{seq}$ to form the initial reasoning state $\mathbf{r}_0$, which initiates the progressive residual reasoning process. This demonstrates that the function $\phi(\cdot)$ exhibits dual role accordingly. When applied to the sequential representation $\mathbf{H}_{seq} \in \mathbb{R}^{s \times d}$ in stage one, it performs behavior-level disentanglement by disentangling mixed patterns across interactions. When applied to the reasoning state $\mathbf{r}_0 \in \mathbb{R}^{d}$ in stage two, it naturally degenerates into a mapping function that projects the account-level state into a latent user representation, transitioning from behaviors to identifying users.

At each step $t$, the reasoning function $\phi(\cdot)$ takes the current state $\mathbf{r}_{(t)}$ and extracts the most dominant undiscovered user:
\begin{equation}
\mathbf{u}_{(t)} = \phi(\mathbf{r}_{(t)}), \quad
\mathbf{r}_{(t+1)} = \mathbf{r}_{(t)} - \mathbf{u}_{(t)},
\end{equation}
where $\mathbf{u}_{(t)}$ denotes the inferred representation of the $t$-th latent user. 
The residual subtraction removes the identified user from the current state, enabling subsequent steps to focus on the remaining unexplained signals. 

The reasoning process terminates adaptively when consecutive inferred users become semantically similar, as this indicates that all latent users have been uncovered. This allows the model to autonomously determine the number of users per account while preserving representational quality and computational efficiency. The similarity is computed as follows:
\begin{equation}
\text{sim}(\mathbf{u}_{(t)}, \mathbf{u}_{(t-1)}) = 
\frac{\mathbf{u}_{(t)} \cdot \mathbf{u}_{(t-1)}}{\|\mathbf{u}_{(t)}\|_2 \, \|\mathbf{u}_{(t-1)}\|_2} > \alpha,
\end{equation}
where $\alpha$ is the hyperparameter to control the depth of reasoning.
Finally, the overall account representation is obtained by aggregating all inferred user embeddings:
\begin{equation}
\mathbf{h}_{final} = \frac{1}{T}\sum_{t=1}^{T} \, \mathbf{u}_{(t)},
\end{equation}
where $T$ means the total reasoning step. 

\subsection{Training Objective}
The final prediction generated by DisenReason is denoted as:
\begin{equation}
P(v_{s+1} \mid S, A) = 
\text{softmax} \left( 
\mathbf{W}_A \cdot [\mathbf{h}_{final}, \mathbf{h}_{{A}}]^{\top} + \mathbf{b}_A 
\right),
\end{equation}
where $\mathbf{h}_{A} \in \mathbf{H}_{{A}}$ is the account representation obtained by Eq.~\ref{Eq.6}.  $\mathbf{W}_A \in \mathbb{R}^{2d \times n}$ is a transformation matrix used to project the prediction results into the dimensional space of candidate items and $\mathbf{b}_A \in \mathbb{R}^{2d}$ is the bias term. The cross-entropy loss function is employed to optimize the DisenReason:
\begin{equation}
\mathcal{L}_{Rec} = - \sum\limits_{k = 1}^m y_{A_k} \cdot \log P(v^{k}_{s+1} \mid S_k, A_k)+(1-y_{A_k}) \cdot \log (1-P(v^{k}_{s+1} \mid S_k, A_k)).
\end{equation}
where $y_{A_k}$ is the one-hot encoding of the ground truth value. $m$ means total number of accounts. To further provide effective optimization signals, we use latent user representations of different reasoning steps as multi-view representations to compute auxiliary recommendation loss. This design encourages the reasoning process to learn meaningful representations for every inferred user. The auxiliary recommendation loss is computed as:
\begin{align}
y_u^{(t)} &= \text{softmax} \left( 
\mathbf{W}_u \cdot [\mathbf{u}_{(t)}, \mathbf{h}_{{A}}] + \mathbf{b}_u 
\right),\\
\mathcal{L}_{Aux} &= -\sum_{k=1}^{m}\sum_{t=1}^{T} y_{A_k} \cdot\log y_{u,k}^{(t)}+(1-y_{A_k}) \cdot\log (1-y_{u,k}^{(t)}),
\end{align}
where $\mathbf{W}_u$ and $\mathbf{b}_u$ are same as $\mathbf{W}_A$ and $\mathbf{b}_A$. For all latent users, we employ shared $\mathbf{W}_u$ and $\mathbf{b}_u$ to reduce the overall number of training parameters.
Then, the overall training objective is defined as:
\begin{equation}
\begin{array}{c}
\mathcal{L}_{all}=\mathcal{L}_{Rec}+\beta\mathcal{L}_{Aux},
\end{array}
\end{equation}
where $\beta$ is the hyperparameter used to control the magnitude of $\mathcal{L}_{Aux}$.

\section{Experiments}
In this section, we first introduce the experimental settings, 
and then analyze the performance of our proposed DisenReason by answering the following \textbf{Research Questions (RQs)}.
\begin{itemize}[label=\textbullet]
\item \textbf{RQ1:} How does our proposed DisenReason perform compared with state-of-the-art baselines?
\item \textbf{RQ2:} What is the contribution of each design in DisenReason?
\item \textbf{RQ3:} How do essential hyperparameters affect DisenReason?
\item \textbf{RQ4:} How robust is DisenReason under different conditions such as sequence length and training data proportion?
\item \textbf{RQ5:} How can DisenReason infer the true number of latent users within shared account?
\end{itemize}

\subsection{Experimental Settings}
\subsubsection{Datasets.}

We conduct experiments on four datasets originally introduced by~\citep{ma2019pi}, namely HvideoE (HV-E), HvideoV (HV-V), HamazonM (HA-M), and HamazonB (HA-B). The first two, HV-E and HV-V, are smart TV datasets that record users’ viewing histories from different TV channels. Specifically, HV-E focuses on logs of educational and instructional videos—covering topics such as sports, health, and medical content. While HV-V contains logs of entertainment content, including television dramas and movies. The remaining datasets, HA-M and HA-B, originate from distinct Amazon domains, corresponding to movie-watching and book-reading activities, respectively.
For the experimental setup, we randomly partition each dataset such that 80\% of the user sequences were used for training and the remaining 20\% for testing. Importantly, for every sequence, the most recent interaction is treated as the ground-truth target item to evaluate model performance. The comprehensive details for the HVIDEO and HAMAZON datasets are detailed in~\Cref{tab:data}.

\begin{table*}[t]
\centering
\caption{Statistical details for HVIDEO and HAMAZON datasets.}
\begin{tabular}{ccccc}
\toprule
\textbf{} & \multicolumn{2}{c}{\textbf{HVIDEO}} & \multicolumn{2}{c}{\textbf{HAMAZON}} \\
\cmidrule(lr){2-3} \cmidrule(lr){4-5}
 & \textbf{E-domain} & \textbf{V-domain} & \textbf{M-domain} & \textbf{B-domain} \\
\midrule
Items & 8,367 & 11,404 & 67,161 & 126,547 \\
Interactions & 2,129,500 & 1,893,784 & 2,196,574 & 2,135,995 \\
Avg. sequence length & 15.85 & 14.09 & 15.27 & 14.84 \\
\midrule
Accounts & \multicolumn{2}{c}{13,714} & \multicolumn{2}{c}{13,724} \\
Sequences & \multicolumn{2}{c}{134,349} & \multicolumn{2}{c}{143,885} \\
Train-sequences & \multicolumn{2}{c}{114,197} & \multicolumn{2}{c}{122,303} \\
Test-sequences & \multicolumn{2}{c}{20,152} & \multicolumn{2}{c}{21,582} \\
\bottomrule
\end{tabular}
\label{tab:data}
\end{table*}

\subsubsection{Evaluation Metrics.}
We adopt two common evaluation metrics to evaluate the model performance, i.e., top-K Recall (RC@K) and top-K Mean Reciprocal Rank (MRR@K), where K = $\{5, 20\}$.


\subsubsection{Implementation Details.}
DisenReason is implemented in PyTorch on NVIDIA RTX 4090 with 24 GB memory. The operating system is Ubuntu
20.04 and the coding platform is Visual Studio Code. We optimize our model with Adam optimizer~\citep{kingma2014adam}. For all datasets, the maximum sequence length is set to 50. To train the DisenReason, we conduct a hyperparameter tuning to identify the optimal hyperparameter configurations, which involves the number of convolutional layers $L\in\{1, 2, 3, 4, 5\}$ in LightGCN, the band-width $B \in \{1, 3, 5, 7, 9\}$ for frequency decomposition,  the termination threshold $\alpha \in \{0.3, 0.4, 0.5, 0.6, 0.7\}$ for progressive reasoning, and the loss weight $\beta \in \{0.1, 0.5, 1.0, 1.5, 2.0\}$
. We use the early stopping strategy to prevent overfitting.

\subsubsection{Baselines.}
To evaluate the effectiveness of DisenReason, we compare it with two categories of baselines: traditional sequential recommendation models and shared-account sequential recommendation models. 

The traditional sequential recommendation baselines include:

\begin{itemize}[label=\textbullet]
\item SASRec~\citep{kang2018self} models the entire user sequence (without any recurrent or convolutional operations), and adaptively considers consumed
 items for prediction.
\item GRURec~\citep{hidasi2015session} introduces GRU into recommendation for the first time. It modified the basic GRU in order to fit the task better by introducing session-parallel mini-batches, mini-batch based output sampling and ranking loss function.
\item  FMLP-Rec~\citep{zhou2022filter} proposes an all-MLP model with learnable filters for sequential recommendation task. The all-MLP architectures endowed our model with lower time complexity, and the learnable filters can be optimized by SGD to adaptively attenuate the noise information in the frequency domain
\item FEARec~\citep{du2023frequency} designs a new frequency-based model to build a hybrid attention manner in both the time and frequency
domains. By working upon defined frequency ramp structure, FEARec employs improved time domain attention to learn both low-high frequency information.
\item BSARec~\citep{shin2024attentive} uses a combination of attentive inductive bias and vanilla self-attention and integrates low and high-frequencies to mitigate oversmoothing.
\item GLINT-RU~\citep{zhang2025glint} combines dense selective GRU with linear complexity, reducing computational costs and accelerating inference speed.
\end{itemize}

The shared-account sequential recommendation baselines include: 
\begin{itemize}[label=\textbullet]
\item $\pi$-Net~\citep{ma2019pi} first introduces the SSR. It improves recommendations performance by leveraging a shared account filter and a
cross-domain transfer mechanism.
\item PSJNet~\citep{sun2021parallel} introduces the ``Split'' and ``Joint'' concepts. ``Split'' is used to identify behavior of different user roles. ``Joint'' is used to discriminate and combine useful user behavior.
\item DA-GCN~\citep{guo2021gcn} models the multiple associations among users and items and the structure information of the transferred knowledge through graph learning.
\item TiDA-GCN~\citep{guo2022time} is an extended version of DA-GCN. To enhance item and account representation learning, it incorporates a time interval-aware message-passing strategy along with an interactive characteristic modeling component, further improving its ability to capture complex dependencies and dynamics within the recommendation process.

\item LightG$\text{C}^2$N~\citep{zhang2025lightweight} designs a convenient and efficient capsule convolutional network to better distinguish the affiliation relationship between items and latent users.

\end{itemize}

\subsection{Overall Performance (RQ1)}

\begin{table*}[tb]
\centering
\small
\caption{Overall performance of different recommendation methods. The best and the second-best performance methods are
denoted in \colorbox{blue!10}{bold} and \colorbox{gray!15}{\underline{underlined}} fonts, respectively. “Improv.” denotes the relative improvement ratios of the proposed approach
over the best performance baselines.}
\begin{adjustbox}{max width=\textwidth} 
\setlength{\tabcolsep}{2.5pt} 
\renewcommand{\arraystretch}{1.05} 
\begin{tabular}{ccccccccccccccccc}
\toprule
\multirow{2}{*}{\textbf{Models}} &
\multicolumn{4}{c}{\textbf{HV\_E}} &
\multicolumn{4}{c}{\textbf{HV\_V}} &
\multicolumn{4}{c}{\textbf{HA\_M}} &
\multicolumn{4}{c}{\textbf{HA\_B}} \\
\cmidrule(lr){2-5} \cmidrule(lr){6-9} \cmidrule(lr){10-13} \cmidrule(lr){14-17}
 & RC@5 & RC@20 & MRR@5 & MRR@20
 & RC@5 & RC@20 & MRR@5 & MRR@20
 & RC@5 & RC@20 & MRR@5 & MRR@20
 & RC@5 & RC@20 & MRR@5 & MRR@20 \\
\midrule
SASRec  & 46.84 & 64.11 & 33.23 & 35.01 & 69.98 & 79.48 & 56.34 & 57.34 & 41.95 & 43.75 & 39.41 & 39.59 & 41.32 & 43.86 & 36.87 & 37.15 \\
GRURec  & 44.61 & 59.64 & 34.84 & 36.33 & 68.36 & 74.93 & 63.19 & 63.85 & \cellcolor{gray!15}\underline{48.08} & \cellcolor{gray!15}\underline{48.75} & 46.87 & 46.94 & 47.88 & 48.52 & 46.44 & 46.51 \\
FMLP-Rec    & 44.27 & 60.77 & 34.43 & 36.09 & 70.23 & 76.45 & 65.01 & 65.64 & 48.02 & 48.35 & \cellcolor{gray!15}\underline{47.51} & \cellcolor{gray!15}\underline{47.53} & \cellcolor{gray!15}\underline{48.62} & \cellcolor{gray!15}\underline{48.97} & \cellcolor{gray!15}\underline{48.01} & \cellcolor{gray!15}\underline{48.04} \\
FEARRec & 50.38 & 64.08 & 40.77 & 42.15 & 70.66 & 77.22 & 64.86 & 65.53 & 40.86 & 43.58 & 37.25 & 37.54 & 39.78 & 42.16 & 36.92 & 37.17 \\
BSARec  & 54.88 & 67.21 & 43.9  & 45.16 & 73.72 & 81.54 & 62.84 & 63.65 & 47.73 & 48.29 & 46.99 & 47.05 & 47.43 & 48.7 & 46.28 & 46.35 \\
GLINT-RU & 44.93 & 60.3  & 34.31 & 35.85 & 70.08 & 77.11 & 63.95 & 65.61 & 43.39 & 47.13 & 45.67 & 45.79 & 47.39 & 47.39 & 47.19 & 47.24 \\
\midrule
$\pi$-Net  & 25.13 & 47.08 & 15.36 & 17.52 & 67.00  & 74.17 & 60.37 & 61.74 & 18.54 & 21.87 & 16.24 & 16.56 & 23.75 & 23.75 & 20.38 & 20.58 \\
PSJNet  & 24.80  & 46.68 & 15.37 & 17.56 & 66.86 & 74.14 & 61.89 & 62.63 & 16.25 & 18.14 & 11.25 & 13.68 & 19.3 & 19.30 & 15.52 & 17.30 \\
DA-GCN  & 51.35 & 66.93 & 35.63 & 37.27 & 75.39 & 82.37 & 59.78 & 60.55 & 22.93 & 23.90 & 20.09 & 20.19 & 23.93 & 24.25 & 21.35 & 21.39 \\
TiDA-GCN & 54.11 & 68.98 & 38.66 & 40.23 & 76.37 & 83.58 & 63.58 & 65.37 & 23.55 & 24.33 & 20.91 & 21.23 & 24.69 & 24.82 & 21.88 & 22.21 \\
LightG$\text{C}^2$N & \cellcolor{gray!15}\underline{61.35} & \cellcolor{gray!15}\underline{72.73} & \cellcolor{gray!15}\underline{46.24} & \cellcolor{gray!15}\underline{47.35} & \cellcolor{gray!15}\underline{79.29} & \cellcolor{gray!15}\underline{84.70} & \cellcolor{gray!15}\underline{66.27} & \cellcolor{gray!15}\underline{66.87} & 46.02 & 48.46 & 41.30 & 41.64 & 47.36 & 48.55 & 44.40 & 44.52 \\
\textbf{Our} & \cellcolor{blue!10}\textbf{63.43} & \cellcolor{blue!10}\textbf{75.01} & \cellcolor{blue!10}\textbf{52.05} & \cellcolor{blue!10}\textbf{53.25} & \cellcolor{blue!10}\textbf{81.89} & \cellcolor{blue!10}\textbf{89.83}  & \cellcolor{blue!10}\textbf{67.89} & \cellcolor{blue!10}\textbf{68.78} & \cellcolor{blue!10}\textbf{49.13} & \cellcolor{blue!10}\textbf{50.05} & \cellcolor{blue!10}\textbf{48.81} & \cellcolor{blue!10}\textbf{48.88} & \cellcolor{blue!10}\textbf{49.56} & \cellcolor{blue!10}\textbf{49.73} & \cellcolor{blue!10}\textbf{49.42} & \cellcolor{blue!10}\textbf{49.44} \\
\midrule
Improv. (\%) & 3.39 & 3.13 & 12.56 & 12.46 & 3.28 & 6.06 & 2.44 & 2.87 & 2.18 & 2.67 & 2.74 & 2.84 & 1.93 & 1.55 & 2.94 & 2.91 \\
\bottomrule
\end{tabular}
\end{adjustbox}
\small
\label{tab:all}
\end{table*}

~\Cref{tab:all} reports the overall performance comparison between our proposed model and various state-of-the-art sequential recommendation (SR) and shared-account sequential recommendation (SSR) baselines across four benchmark datasets.  DisenReason achieves the best performance across all four datasets. Specifically, on the HV-E and HV-V datasets, DisenReason outperforms the strongest SSR baseline, LightGC$^2$N, with a relative improvement ranging from 2.06\% to 12.47\%. Notably, LightGC$^2$N outperforms SR baselines such as FEARec, BSARec, and GLINT-RU. This superiority highlights the advantage of explicitly modeling latent user interactions under shared-account scenarios. SR models assume that all behaviors in a sequence originate from a single user and thus fail to capture the heterogeneous preferences within an account, LightGC$^2$N leverages graph-based message passing and attention mechanisms to propagate user-specific information, enabling more accurate representation learning. 
DisenReason outperforms LightGC$^2$N, which indicates that learning the actual number of real users in each account, rather than relying on a predefined number, provides a helpful performance boost.

On the HA-M and HA-B datasets, DisenReason also has a relative improvement ranging form 1.93\% to 2.94\%. Interestingly, LightGC$^2$N as the best-performing SSR method, does not surpass traditional SR models on the HA-M and HA-B datasets. In contrast, SR methods such as GRURec and FEARec even outperform LightGC$^2$N to some extent. Upon deeper analysis, we found that HV-V and HV-E are real shared-account datasets, whereas HA-M and HA-B are synthetic ones. Therefore, we hypothesize  that this performance discrepancy may be attributed to the fundamental difference between real and synthetic shared-account behaviors.

In summary, the overall results confirm that DisenReason enables a more faithful and interpretable representation of shared-account behaviors on real and synthetic shared-account scenario, leading to substantial improvements in recommendation performance.

\subsection{Ablation Study (RQ2)}

\begin{table*}[t]
\centering
\caption{Ablation study of DisenReason on four datasets.
‘w/o’ denotes the model variant without the corresponding
design.}
\small
\begin{tabular}{ccccccccc}
\toprule
\multirow{2}{*}{\textbf{Models}} & 
\multicolumn{2}{c}{\textbf{HV-E}} & 
\multicolumn{2}{c}{\textbf{HV-V}} & 
\multicolumn{2}{c}{\textbf{HA-M}} & 
\multicolumn{2}{c}{\textbf{HA-B}} \\
\cmidrule(lr){2-3} \cmidrule(lr){4-5} \cmidrule(lr){6-7} \cmidrule(lr){8-9}
 & RC@5 & MRR@5 & RC@5 & MRR@5 & RC@5 & MRR@5 & RC@5 & MRR@5 \\
\midrule
w/o LightGCN        & 61.96 & 51.32 & 79.84 & 67.68 & 48.61 & 48.52 & 49.44 & 49.35 \\
w/o Behavior Disentanglement & 55.09 & 33.35 & 77.81 & 53.17 & 35.89 & 22.62 & 36.21 & 22.67 \\
w/o Adaptive Fusion       & 62.51 & 50.88 & 81.59 & 67.64 & 49.05 & 48.76 & 49.54 & 49.39 \\
w/o Residual Operation       & 63.32 & 51.37 & 81.34 & 67.69 & 49.08 & \cellcolor{blue!10}\textbf{48.84} & 49.50 & 49.32 \\
\midrule
\textbf{Our} & \cellcolor{blue!10}\textbf{63.43} & \cellcolor{blue!10}\textbf{52.05} & \cellcolor{blue!10}\textbf{81.89} & \cellcolor{blue!10}\textbf{67.89} & \cellcolor{blue!10}\textbf{49.13} & 48.81 & \cellcolor{blue!10}\textbf{49.56} & \cellcolor{blue!10}\textbf{49.39} \\
\bottomrule
\end{tabular}
\label{tab:ab}
\end{table*}

In this section, we conduct an ablation study to examine the contribution of each key designs in DisenReason, including representation construct (w/o LightGCN), frequency disentanglement operation (w/o Behavior Disentanglement), adaptive fusion operation for different behavioral patterns (w/o Adaptive Fusion), and residual operation for reasoning (w/o Residual Operation).~\Cref{tab:ab} reports the performance of different variants across the four datasets. The experimental results are analyzed as follows:
\begin{itemize}[label=\textbullet]
   \item \textbf{Effect of LightGCN.} When removing the LightGCN, the overall performance decreases noticeably, especially on the HV-E and HV-V. This degradation indicates that LightGCN plays a crucial role in capturing global account–item collaborative signals, which provide a stable semantic foundation for the subsequent modules.

   \item \textbf{Effect of Behavior Disentanglement.} In the first stage, we do not perform behavior disentanglement and directly adopt the last item of the account sequence as the reasoning pivot. Removing the behavior disentanglement causes the most significant performance drop among all variants. For instance, MRR@5 decreases from 52.05 to 33.35 on HV-E and from 67.89 to 53.17 on HV-V. This dramatic decline demonstrates the effectiveness of obtaining a pivot that represents the entire account, rather than directly using the last item in the account as the pivot. Directly using the last item in the account as the pivot for user reasoning leads to an incomplete understanding of the latent behaviors within the account during the subsequent reasoning process. In such cases, progressive reasoning cannot effectively infer the hidden users behind the account, which in turn degrades the recommendation performance.

   \item \textbf{Effect of Adaptive Fusion.}
The adaptive fusion enables the identification of the importance of each latent behavior, thereby combining them into a more realistic account-level representation. When this operation is removed, all metrics show a decline, demonstrating that considering the importance of different behaviors within the account helps in learning a better account behavior representation.

   \item \textbf{Effect of Residual Operation.} The residual operation prevents redundant semantic interpretation and allow subsequent steps to focus on unexplained user in reasoning processing. Removing it results in only a slight performance decline. This is because the residual operation plays a semantic refinement role, which polishes latent user semantic boundaries. Consequently, removing it yields only a moderate performance drop.
\end{itemize}

\subsection{Hyperparameter Sensitivity (RQ3)}

\begin{figure*}
  \centering
  \includegraphics[width=0.8\linewidth]{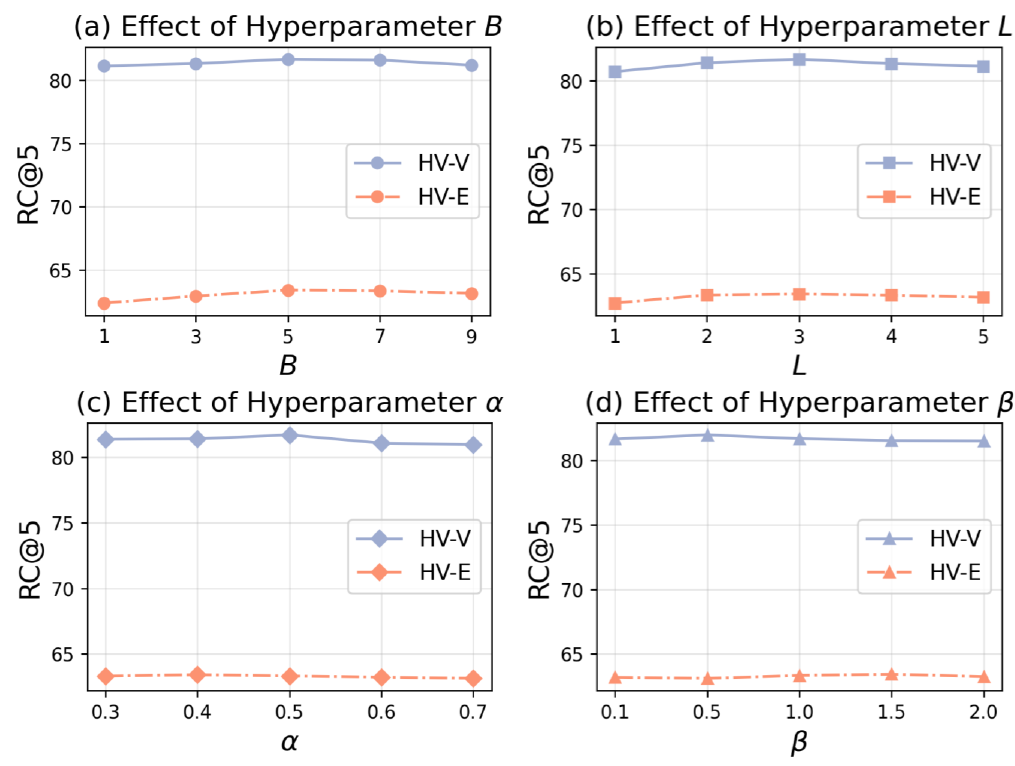} 
  \caption{Hyperparameter analysis of DisenReason on HV-V and HV-E dataset.}
  \label{fig:para}
\end{figure*}

\begin{figure*}
  \centering
  \includegraphics[width=0.8\linewidth]{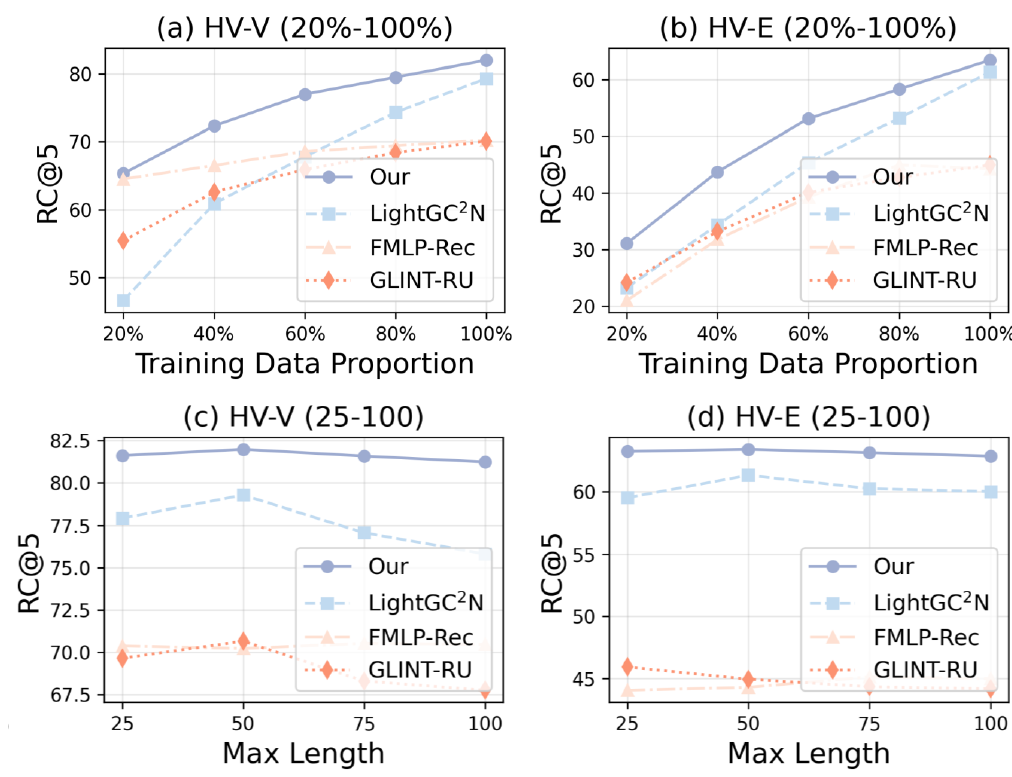} 
  \caption{Performance comparison of DisenReason and baselines under different sequence lengths and training data scales on HV-V and HV-E datasets.}
  \label{fig:robust}
\end{figure*}

To evaluate the sensitivity of DisenReason with respect to different hyperparameter settings, we conduct sensitivity experiments on HV-V and HV-E, as illustrated in~\Cref{fig:para}. The analysis is as follows: 
\begin{itemize}[label=\textbullet]
 
   \item \textbf{Hyperparameter $B$.} 
$B$ is used to adjust the strength of frequency disentanglement. As illustrated in~\Cref{fig:para}(a), the model achieves optimal performance when $B=5$. $B$ is too small may fail to properly disentangle the complete behavioral patterns, leading to fragmentation of these patterns. In contrast, an excessively large $B$ may result in behavioral patterns that remain entangled. Thus, a moderate $B$ achieves the best balance between disentanglement granularity and representation stability.

   \item \textbf{Hyperparameter $L$.} 
$L$ controls the convolution depth of LightGCN. From ~\Cref{fig:para}(b), the model performance improves with deeper propagation up to $L=3$, then slightly decreases. 
This indicates that deeper layers enhance the ability to capture collaborative relations but may introduce high-order noise, resulting in over-smoothing. Therefore, $L=3$ provides a good trade-off between information aggregation and model generalization.

   \item \textbf{Hyperparameter $\alpha$.} 
$\alpha$ serves as the termination signal during the user inference process. As shown in~\Cref{fig:para}(c), the proposed model achieves the best performance when $\alpha=0.5$ and $\alpha=0.4$ on HV-V and HV-E, respectively . 
The threshold $\alpha$ determines when the reasoning process stops based on the similarity between consecutive inferred users. An appropriate $\alpha$ prevents redundant user generation.

   \item \textbf{Hyperparameter $\beta$.} 
$\beta$ controls the strength of the supervision signal.
As illustrated in~\Cref{fig:para}(d), when the value of $\beta$ is set to 1.0 and 0.5, DisenReason achieves optimal performance on the HV-V and HV-E datasets, respectively. This indicates that providing additional supervisory signals during the progressive reasoning process (stage two) helps enhance the diversity of the reasoning process, which in turn improves recommendation performance.
\end{itemize}

\subsection{Robustness Analysis (RQ4)}
To assess the robustness of DisenReason, we conduct two sets of experiments focusing on sequence length and training data. The first experiment varied the maximum sequence length, while the second controlled the proportion of available training data. The results, as shown in~\Cref{fig:robust}, consistently demonstrate the superior stability and robustness of DisenReason.

When varying the maximum sequence length from 25 to 100, DisenReason consistently outperforms all baselines, including LightGC$^2$N, FMLP-Rec, and GLINT-RU, across both the HV-V and HV-E datasets. Unlike the baselines, whose performance fluctuates or even deteriorates as the sequence becomes longer due to the accumulation of noisy or redundant interactions, DisenReason maintains a stable performance. Under low resource conditions, where the amount of training data ranges from 20\% to 100\%, DisenReason again exhibits remarkable resilience and data efficiency. Even when trained with only 20\% of the data, our model surpasses the strongest baseline. More importantly, DisenReason demonstrates a smooth and nearly linear performance improvement as the data scale increases compared to baselines. 

\begin{figure*}
  \centering
  \includegraphics[width=\linewidth]{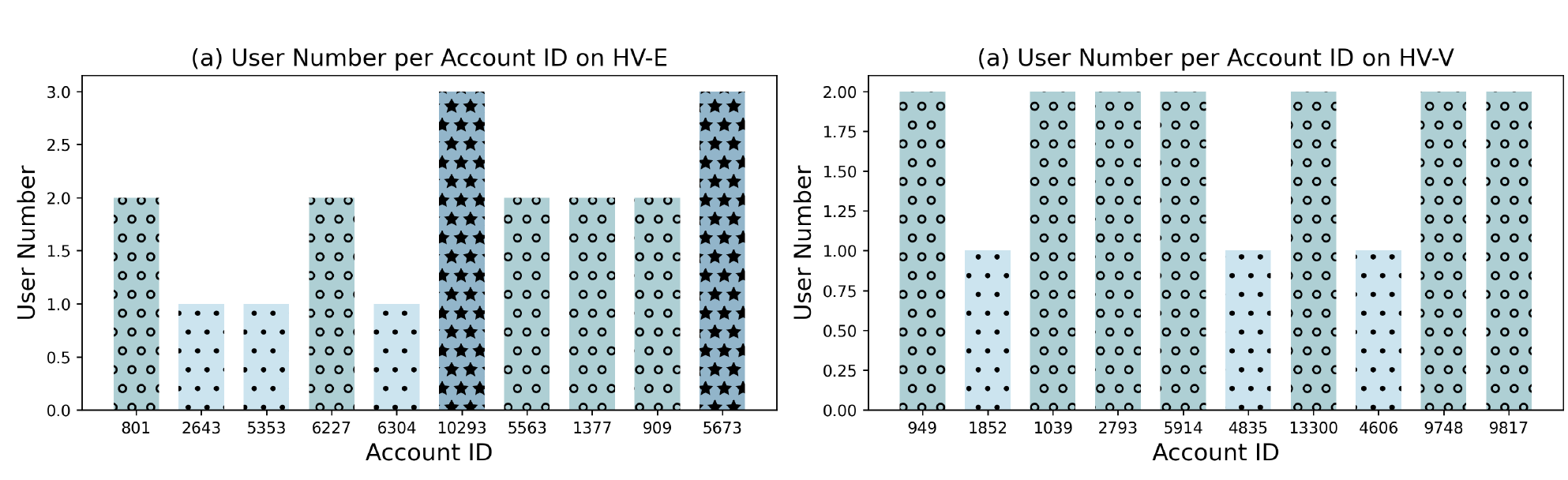} 
  \caption{Latent user number per account ID on HV-E and HV-V dataset. It illustrates the number of latent users inferred for each account ID through the reasoning process.}
  \label{fig:user}
\end{figure*}

\subsection{In-depth Analysis (RQ5)}
In this section, we present a detailed analysis of 10 accounts selected from the HV-E and HV-V datasets and visualize their inferred latent user counts, as shown in~\Cref{fig:user}. The experiment highlights the effectiveness of the reasoning mechanism in adaptively inferring the number of latent users across different accounts. The results demonstrate that DisenReason can dynamically adjust to varying complexities in shared-account scenarios. Specifically, on the HV-E dataset, some accounts such as those with IDs 10293 and 5673, exhibit higher latent user counts, while others, like 6304 and 2643, show fewer latent users. This diversity in inferred user counts reflects the model's ability to distinguish between accounts with different interaction patterns and complexities. This phenomenon also occurs in the HV-V dataset.

The key insight from this experiment is that the two-stage reasoning process in DisenReason effectively infers the number of latent users, allowing it to adapt to a wide range of real-world scenarios. In many shared account contexts, the number of users per account is not predefined and can vary significantly. This adaptability is crucial for accurately modeling and predicting user behaviors in situations where the structure of shared accounts is not fixed. Through this experiment, we validate the DisenReason's capacity to handle the inherent variability in user counts and interactions, showcasing its robustness in real-world shared-account environments.

\section{Theoretical and Practical Implications}
Theoretically, our proposed DisenReason addresses critical limitation of existing shared-account sequential recommendation (SSR) methods, the unrealistic assumption of a fixed number of latent users per account, by introducing a novel two-stage reasoning framework. In the first stage, Fast Fourier Transform is applied to disentangle the mixed sequence in the frequency domain, obtaining distinct behavioral patterns within the account, which are then adaptively fused via a Mixture-of-Experts network to derive an account-level reasoning pivot. In the second stage, we propose a progressive residual reasoning mechanism that iteratively extracts latent users from the account-level pivot and adaptively terminates based on semantic similarity. These designs collectively enable DisenReason to make meaningful contributions to the SSR literature, particularly in advancing reasoning-enhanced recommendation for shared-account scenarios.

Practically, DisenReason demonstrates clear utility in real-world settings where account sharing is prevalent. Extensive experiments across four benchmark datasets, including both real-world and synthetic shared-account scenarios, show consistent and substantial performance gains over state-of-the-art baselines, with relative improvements reaching up to 12.56\% in MRR@5. The framework proves particularly effective in capturing the dynamic and variable number of users behind each account, a common challenge in commercial platforms such as streaming services, e-commerce sites, and smart home systems. By adaptively inferring latent user counts and learning account-level representations, DisenReason enables more personalized and context-aware recommendations without requiring explicit user identification or manual configuration. This work offers actionable insights for practitioners aiming to build scalable, robust, and user-centric recommendation services that accommodate the realities of shared account usage.

\section{Conclusion and Future Work}
In this paper, we address the core challenge of Shared-Account Sequential Recommendation (SSR): how to accurately infer and represent the dynamic number and behavior patterns of latent users behind a shared account. Existing approaches often assume a fixed number of latent users, which is unrealistic given the varying number of users sharing an account in practice. To overcome this, we propose DisenReason. It operates in two stages: (1) Behavior Disentanglement for Pivot, where we establish a unified pivot reflecting the collective dynamics of the shared account, and (2) Progressive Residual Reasoning for Latent User, where we progressively infer latent users from this pivot. This two-stage reasoning effectively uncover the number of latent users, leading to more precise and personalized recommendations.
Our empirical experiments validate the effectiveness of this framework, outperforming existing methods on multiple datasets.

Despite its promising performance, DisenReason suffers from fixed bandwidth segmentation that may overlook fine-grained behavioral boundaries and parameter coupling between two stages that blurs task objectives. Future work will investigate adaptive frequency decomposition to better preserve behavioral integrity, and develop decoupled architectures or meta-learning mechanisms to distinguish global aggregation from personalized reasoning. These directions aim to improve both the accuracy and interpretability of latent user inference in shared account scenarios.
\printcredits

\section*{Acknowledgements}
 This paper was supported by the National Natural Science Foundation of China (Grant Nos. 62176028).

\section*{Declaration of competing interest}
 The authors declare that they have no competing interests.

 \section*{Data availability}
  Data will be made available on request.
 







\bibliographystyle{cas-model2-names}

\bibliography{ref}



\end{document}